\documentclass[twocolumn]{aastex63}
\usepackage{hyperref}
\usepackage{amsmath,amssymb,mathrsfs}
\usepackage{soul}
\usepackage{lineno}

\def\ztfl {ZTF\,J213056.71+442046.5}
\def\ztf {ZTF\,J2130}

\def\hd {HD\,49798}

\def\cd {CD\, --30$^\circ$\,11223}
\def \xmm {\emph{XMM-Newton}}

\def \mdot {\dot M}

\def\msun{{\rm M}_{\odot}}
\def\rsun{R_{\odot}}
\def\lsun{L_{\odot}}

\def\mdotn {\dot M_{\rm -9}}

\def\ltsima{$\; \buildrel < \over \sim \;$}
\def\lsim{\lower.5ex\hbox{\ltsima}}
\def\gtsima{$\; \buildrel > \over \sim \;$}
\def\gsim{\lower.5ex\hbox{\gtsima}}
\def\nh{$N_{\rm H}$}

\graphicspath{{fig/}}

\defcitealias{Davies81}{DP81} 

\received{March 28, 2022 }
\revised{ April 18, 2022}
\accepted{April 19, 2022}
\submitjournal{ApJ}

\shorttitle{\xmm\ observation of \ztf\ }
\shortauthors{Mereghetti et al.}

\begin{document}

\title{X-ray observation of the Roche-lobe filling white dwarf plus hot subdwarf system \ztfl\ }

\correspondingauthor{Sandro Mereghetti}
\email{sandro.mereghetti@inaf.it}

\author[0000-0003-3259-7801]{S.~Mereghetti}
\affiliation{INAF -- Istituto di Astrofisica Spaziale e Fisica Cosmica, Via A. Corti 12, I-20133 Milano, Italy}
 
\author[0000-0000-0000-0000]{N.~La Palombara}
\affiliation{INAF -- Istituto di Astrofisica Spaziale e Fisica Cosmica, Via A. Corti 12, I-20133 Milano, Italy}

\author[0000-0002-6540-1484]{T.~Kupfer}
\affiliation{Department of Physics and Astronomy, Texas Tech University, PO Box 41051, Lubbock, TX 79409, USA}

\author[0000-0002-2498-7589]{T.R.~Marsh}
\affiliation{Department of Physics, University of Warwick, Gibbet Hill Road, Coventry CV4 7AL, UK}

\author[0000-0001-7983-8698]{C.M.~Copperwheat }
\affiliation{Astrophysics Research Institute, Liverpool John Moores University, IC2, Liverpool Science Park, L3 5RF, UK}

\author[0000-0001-5253-3480]{K.~Deshmukh}
\affiliation{Department of Physics and Astronomy, Texas Tech University, PO Box 41051, Lubbock, TX 79409, USA}

\author[0000-0003-4849-5092]{P.~Esposito}
\affiliation{Scuola Universitaria Superiore IUSS Pavia, Palazzo del Broletto, piazza della Vittoria 15, 27100 Pavia, Italy}
\affiliation{INAF -- Istituto di Astrofisica Spaziale e Fisica Cosmica, Via A. Corti 12, I-20133 Milano, Italy}

\author[0000-0003-0976-4755]{T.~Maccarone}
\affiliation{Department of Physics and Astronomy, Texas Tech University, PO Box 41051, Lubbock, TX 79409, USA}

\author[0000-0002-3869-2925]{F.~Pintore}
\affiliation{INAF -- IASF Palermo, via U. La Malfa 153, 90146 Palermo,   Italy}

\author[0000-0001-6641-5450]{M.~Rigoselli}
\affiliation{INAF -- Istituto di Astrofisica Spaziale e Fisica Cosmica, Via A. Corti 12, I-20133 Milano, Italy}
   
\author[0000-0002-9396-7215]{L.~Rivera Sandoval}
\affiliation{Department of Physics and Astronomy, University of Texas Rio Grande Valley, Brownsville, TX 78520, USA}
  
\author[0000-0002-6038-1090]{A.~Tiengo}
\affiliation{Scuola Universitaria Superiore IUSS Pavia, Palazzo del Broletto, piazza della Vittoria 15, 27100 Pavia, Italy}
\affiliation{INAF -- Istituto di Astrofisica Spaziale e Fisica Cosmica, Via A. Corti 12, I-20133 Milano, Italy}
\affiliation{Istituto Nazionale di Fisica Nucleare, Sezione di Pavia, via A. Bassi 6, I-27100 Pavia, Italy}

\begin{abstract}

 \ztfl\ is the prototype of a small class of recently discovered compact binaries  composed of a white dwarf and a hot subdwarf that fills its Roche-lobe. Its orbital period of only 39 min is the shortest known  for the objects in this class. Evidence for a high orbital inclination ($i$=86$^{\circ}$) and for the presence of an accretion disk has been inferred from a detailed modeling of its optical photometric and spectroscopic data. We report the results of an \xmm\ observation carried out on 2021 January 7. \ztfl\ was clearly detected by the Optical Monitor, which showed a periodic variability in the UV band (200-400 nm), with a light curve similar to that seen at longer wavelengths. Despite accretion on the white dwarf at an estimated rate of the order of 10$^{-9}$  $\msun$ yr$^{-1}$, no X-rays were detected with the EPIC instrument, with a limit of $\sim$10$^{30}$ erg s$^{-1}$ on the 0.2-12 keV luminosity. 
We discuss possible explanations for the lack of a strong X-ray emission from this system.
\end{abstract}



\section{Introduction}
\label{sec:intro}
  
A   class of close binary systems consisting of Roche-lobe filling hot subdwarfs with white dwarf (WD) companions has been recently identified thanks to   optical surveys devoted to the discovery of transient and variable sources \citep{kup20b}. 
Hot subdwarfs (see \citealt{heb16}, for a review)  are low mass stars ($\sim$0.3--0.8 $\msun$), spectrally classified as sdO or sdB based on their effective temperature above or below 38 kK, respectively.  
They are believed to originate from the evolution of stars that have lost most of their envelopes during the red-giant phase. 
One of the  possible mechanisms responsible for the loss of the massive hydrogen envelopes necessary to form hot subdwarfs, is non-conservative mass transfer in a binary.  This is observationally supported by the large number of hot subdwarfs, particularly of sdB type, found in close binary systems \citep{mor03,nap04}.
Evolutionary models  predict that most of the companions of sdB stars in systems with orbital periods shorter than $\sim$10 days should be either late type main sequence stars or WDs \citep{han02,han03}.

Hot subdwarfs with WD companions are particularly interesting  because  they might be among the progenitors of type Ia supernovae \citep{ibe94}  and some of the nearby systems with ultrashort orbital period are expected to be sources of gravitational waves  detectable by LISA \citep{kup18}.
Furthermore, the determination of the mass and evolutionary stage of the two components, as done, e.g., for KPD 0422+5421 \citep{oro99}, KPD 1930+2752 \citep{gei07},  and \hd\ \citep{mer09}, can shed light on the poorly known processes that take place during the common-envelope evolutionary phase.

The hot subdwarfs with WD companions that exit the common-envelope phase at orbital periods $P_{\rm orb}\lsim2$ hr will reach contact while the sdB is still burning He \citep{bau21}. Due to the emission of gravitational waves, the orbit will shrink until the sdB star fills its Roche-lobe at an orbital period  dependent on the evolutionary stage of the subdwarf 
\citep{ibe91,yun08}.
The known population of sdB binaries consists mostly of systems with orbital periods too large to start accretion before the sdB turns into a WD \citep{kup15}.
Until recently, only five binaries of this kind with  $P_{\rm orb}$\lsim\ 2 hr were known \citep{ven12,kup17a,kup17b,pel21,kup22}.
The sdBs in these systems do not fill their  Roche-lobe.  X-ray observations of one of them (\cd , $P_{\rm orb}$=70 min) provided an upper limit on the X-ray luminosity indicating a wind mass-loss  rate  from the sdB star lower than  $\sim3\times10^{-13}$ $\msun$ yr$^{-1}$  \citep{mer14}.

The first Roche-lobe filling subdwarf with a WD companion, \ztfl\ (hereafter \ztf ,  \citealt{kup20}) was identified in a search for periodic objects in a  catalog of $\sim$ 40,000 hot subdwarf candidates  \citep{gei19} and first discovered in a dedicated high-cadence survey  with the Zwicky Transient Facility \citep{bel19}.   
\ztf\ consists of a sdB star  with mass $\sim0.35 \msun$ and radius 0.13$\pm$0.1 $\rsun$ and  a  WD of $\sim0.56 \msun$.  The orbital period of this system, $P_{\rm orb}$ = 39 min (also independently discovered by G.~Murawski\footnote{https://www.aavso.org/vsx/index.php?view=detail.top\&oid=689728}),  is the shortest of any known hot subdwarf binary. 
The radial velocity curve of \ztf\ indicates a circular orbit with  a velocity semi-amplitude $K$ = 429 $\pm$ 28 km s$^{-1}$.  The periodic variability is primarily caused by the ellipsoidal modulation of the subdwarf, due to its tidal deformation under the gravitational influence of the  companion. However, the optical light-curve cannot be described with a simple model composed of two detached stars.   

\citet{kup20} showed that the subdwarf in this system fills its Roche-lobe and the WD is surrounded by an optically thick accretion disk of size $\sim0.14 \rsun$.  The formation of a disk is not surprising in a Roche-lobe filling system, and its presence suggests the  possible emission of accretion-powered  X-rays.  A short pointing (1 ks) carried out with the X-ray Telescope on the \emph{Neil Gehrels Swift Observatory} in 2019 did not detect X-rays \citep{riv19}, giving a flux upper limit of the order of a few 10$^{-13}$ erg cm$^{-2}$ s$^{-1}$.  Here we report on a much deeper observation of \ztf\ obtained with the \xmm\ satellite.
In the following we use the source distance of 1.29$^{+0.06}_{-0.04}$ kpc, derived from  the Gaia EDR3 parallax  \citep{bai21}.

\section{Data analysis and results} 
\label{sec:data analysis}

The field of \ztf\ was observed  with  the \xmm\ satellite for about 65 ks starting at 23:47 UT of 2021 January 6.
We used the X-ray data obtained with the EPIC instrument, comprising one pn \citep{str01} and two  MOS cameras \citep{tur01} operating in the 0.2--12 keV energy range,  and with the Optical Monitor (OM,   \citealt{mas01}). 
The three EPIC cameras were used in full frame mode  with the thin optical filter. The OM provided data in the  UVW1 band ($\lambda_{eff}$=291 nm, $\Delta\lambda$= 83 nm) for the first 26.4 ks and in the UVW2 band ($\lambda_{eff}$=212 nm, $\Delta\lambda$=50 nm ) for the remaining part of the observation.  
All the data were processed with Version 19.0.0 of the Science Analysis Software (\textsc{sas})\footnote{https://xmm-tools.cosmos.esa.int/external/xmm\_user\_support/ documentation/sas\_usg/USG/}.

\subsection{  EPIC X-ray data}
\label{sec-epic}

The EPIC data were affected by periods of high background caused by  soft proton flares. Removing the corresponding time intervals 
resulted  in  net exposure times  of 31 ks  for the pn and 37 ks for the MOS. We extracted images from the three cameras in five different energy ranges (0.2-0.5, 0.5-1, 1-2, 2-4.5, and 4.5-12 keV), both using the  cleaned data and the whole observation. Many sources were clearly visible in these images, but none  at the  position of \ztf . 

We performed a more detailed analysis by  means of a source  detection procedure over the whole field of view using the data of the three cameras simultaneously.  This was done in the five energy bands given above using the SAS task \textsc{edetect\_chain}\footnote{https://xmm-tools.cosmos.esa.int/external/sas/current/ doc/edetect\_chain/index.html}, 
as described in the Appendix, and resulted in the detection of  51 sources (see Table~\ref{table-sources}) 
above a threshold likelihood $L$ = 10, corresponding to a confidence level of 99.9955 \%.  
 \ztf\ was not detected and we derived  an upper limit of 1.68 counts ks$^{-1}$ on its count rate in the 0.2-12 keV energy range (with a threshold likelihood $L$ = 8,  i.e. 99.97 \% confidence level).  The limits in the 0.2-0.5 and 0.5-4.5 keV ranges were 0.69 and 1.34 counts ks$^{-1}$, respectively.

\subsection{ OM ultraviolet  data}

\begin{figure}
    \centering
    \includegraphics[width=0.52\textwidth]{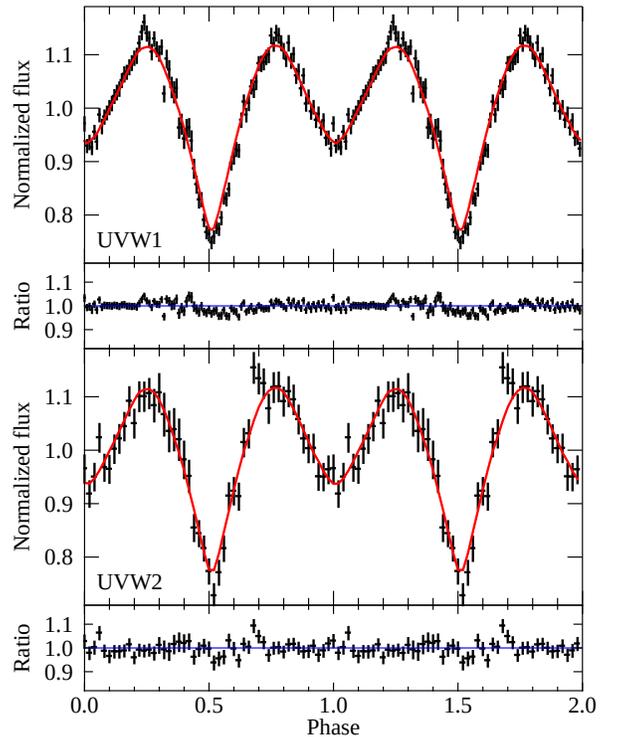}
    \caption{The black points show the Optical Monitor light curves of \ztf\ in two UV bands folded at the orbital period. The red  lines show for comparison the  g-band light curves.
    The second and fourth panels from top show the ratios between the UV and g-band light curves.
    The WD is between the observer and the sdB star at phase 0.5.}
    \label{fig-lc}
\end{figure}

\ztf\ was clearly detected by the OM, with average count rates of 22 counts s$^{-1}$ in the UVW1  filter  (291 nm) and 2.5 counts s$^{-1}$ in the UVW2 filter (212 nm) (Table \ref{table:1}).   Figure~\ref{fig-lc} shows the OM light curves in the two filters folded at the orbital period according to the ephemeris of \citet{kup20}.
In both filters, the orbital modulation is consistent with that seen in the optical band, as it is shown by the superimposed solid lines representing the profiles obtained in the g-band with HiPERCAM \citep{kup20}.  Some small residuals remain.  These are most likely to be indicating factor of $\sim2$ variability in the accretion disc flux.

A search of periodicities was performed using the OM-UVW1 data, but the periodogram did not reveal any additional periodicity beyond the known orbital period of 39 min.

\begin{table*}
	\centering
	\caption{Ultraviolet and X-ray measurements of \ztf\ obtained with \xmm . Errors in luminosity include the error in distance and are given at $1\sigma$. Magnitudes are not dereddened. The upper limit in the X-ray luminosity is valid for the blackbody and thermal bremsstrahlung models.
	}
	\label{table:1}
 \begin{tabular}{l c c c  c c r}
		\hline
		Instrument & Band & $\lambda_{eff}$ & Exposure & Mean Magnitude & Flux & Luminosity\\
		           &      & (nm)           &(ks)       & (Vega)    & $\times 10^{-11}$ erg cm$^{-2}$ s$^{-1}$ & $\times10^{33}$ erg s$^{-1}$ \\
		\hline
		OM & $UVW1$     & 291  &   26.4 & $13.849 \pm 0.289$ & $3.16\pm0.84$    & $6.3\pm1.4$\\ 
		OM & $UVW2$     & 212	  & 37.32   & $13.809 \pm 0.003$ & $3.27\pm0.01$ & $6.51\pm 0.02$ \\
		EPIC   & 0.2-12 keV & --&31 (pn), 37 (MOS) &  --  & $<0.0012$ &  $<0.0025$\\
		\hline
	\end{tabular}
\end{table*}

\begin{figure}
    \centering
    \includegraphics[width=0.45\textwidth]{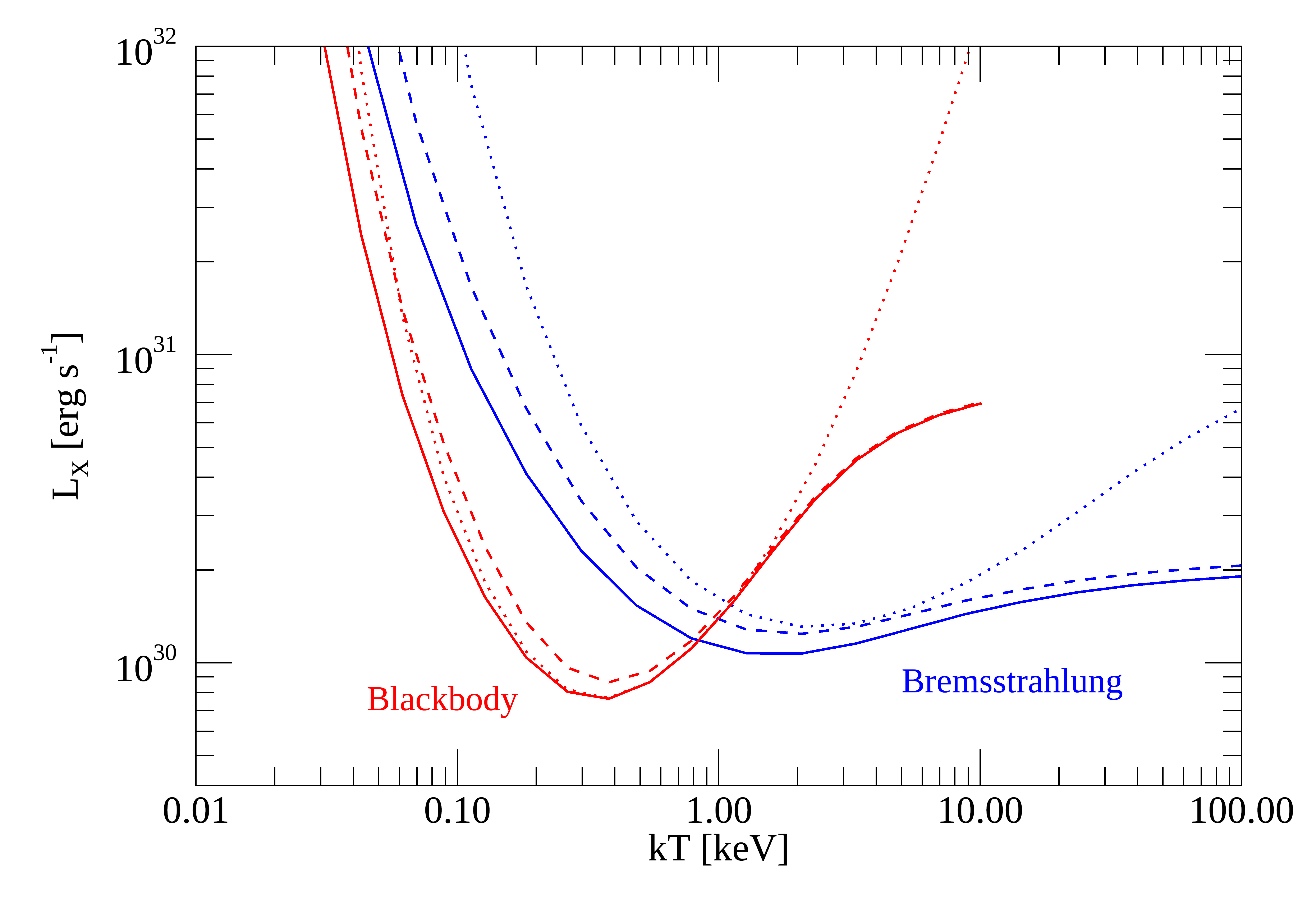}
     \caption{Upper limits on the luminosity of \ztf\ as a function of temperature, for an assumed blackbody (red) or thermal bremsstrahlung (blue) spectrum. The solid and dashed lines give the unabsorbed luminosity in the 0.2-12 keV range for \nh = 1.7$\times10^{21}$ cm$^{-2}$ and 2.4$\times10^{21}$ cm$^{-2}$, respectively. The dotted lines give the bolometric (0.001-100 keV) unabsorbed luminosity for  \nh = 1.7$\times10^{21}$ cm$^{-2}$.  }
    \label{fig-ul}
\end{figure}

\section{Discussion} \label{sec:discussion}
 
Given the well known distance of \ztf , the most important factors affecting the conversion of the count rate upper limits derived in Section~\ref{sec-epic} to an upper limit on the luminosity are the spectral shape assumed for the X-ray emission  and  the amount of interstellar absorption.  
Accreting  WDs have complex X-ray spectra, with different components depending on many   factors, including,  e.g., the WD mass and magnetic field, the geometry of the accretion flow, the binary orientation wrt the line of sight (see, e.g., \citealt{muk17}, for a review). Optically thin thermal emission originates from the shocked plasma, which in magnetic WDs is usually channeled in an accretion column. Depending on the plasma temperature, significant line emission can be present, in addition to the continuum bremsstrahlung flux.  Part of the emission from the accretion column can heat the star surface and will then be re-radiated as an optically thick soft thermal component.  Other spectral components can arise from Compton reflection of the hardest X-rays,  from the inner part of an accretion disk,   from the boundary layer, and from a hot corona above the disk.  Detailed models for many of these emission components have been developed, but for our purpose it is sufficient to adopt two simple models, blackbody and thermal bremsstrahlung,   which provide an adequate phenomenological description of the hard and soft components observed in accreting WDs. 

 The 3-D extinction maps of \citet{gre19} give a reddening of $E(g-r)$=0.18 at the distance of \ztf . This corresponds to  $A_V\sim$0.6 mag, which, using the   relation between $A_V$ and  interstellar absorption   derived by \citet{bah15},   gives   \nh=1.7$\times10^{21}$ cm$^{-2}$.
This value is consistent with the  upper limit provided by the total column density in this direction, \nh=2.4$\times$10$^{21}$ cm$^{-2}$ \citep{HI4PI}.

Therefore, we computed the count rate to flux conversion factors for blackbody and   thermal bremsstrahlung models and two representative values of absorption (the nominal and the maximum values,  1.7 and 2.4 $\times10^{21}$ cm$^{-2}$, respectively). This was done using the response matrix for the EPIC instrument (i.e. the sum of the pn and two MOS) computed for the position of \ztf\  in the instrument field of view. 
The resulting upper limits on the luminosity are plotted as a function of the spectral temperature in Fig.~\ref{fig-ul}.  Except for the case of very soft spectral shapes, for which most of the emission falls outside the energy range sampled by the EPIC instrument, the limits are quite constraining and indicate a luminosity  below  $\sim10^{30}$ erg s$^{-1}$ for a wide range of assumed temperature values. 

 By modeling the evolutionary history of \ztf , \citet{kup20} estimated that the WD in this system is accreting at a rate of  $\sim10^{-9}$ $\msun$ yr$^{-1}$. Taken at face value, this would imply an accretion-powered luminosity of

\begin{equation}
L_{acc} =  G  \frac{M}{R}  \mdot = 5\times10^{33}  \mdotn~~{\rm  erg~s^{-1}},
\end{equation}
  
\noindent
where G is the gravitational constant,  $\mdotn$ is the mass accretion rate in units of 10$^{-9}$ $\msun$ yr$^{-1}$, and we  used the   values of $M$=0.545 $\msun$ and $R$=9400 km appropriate for this WD.
This luminosity is orders of magnitude larger than the upper limit derived with \xmm , but there are possible explanations for the lack of detectable X-ray emission in \ztf .  

First, it must be considered that  about half of the accretion power is  dissipated in the accretion disk, which is too cold to emit in the X-ray band. Most of the disk luminosity occurs at IR/optical/UV wavelengths and  is outshined by the much brighter emission from the hot subdwarf present in this system. The latter has an effective temperature of 42,400 K and a bolometric luminosity of  41$\pm$9  $\lsun$. 
From the absence of optical emission lines, \citet{kup20} concluded that the accretion disk contributes at most 3\% of the overall luminosity, which is consistent with the above estimate of L$_{acc}$.  

\begin{figure}
    \centering
    \includegraphics[width=0.5\textwidth]{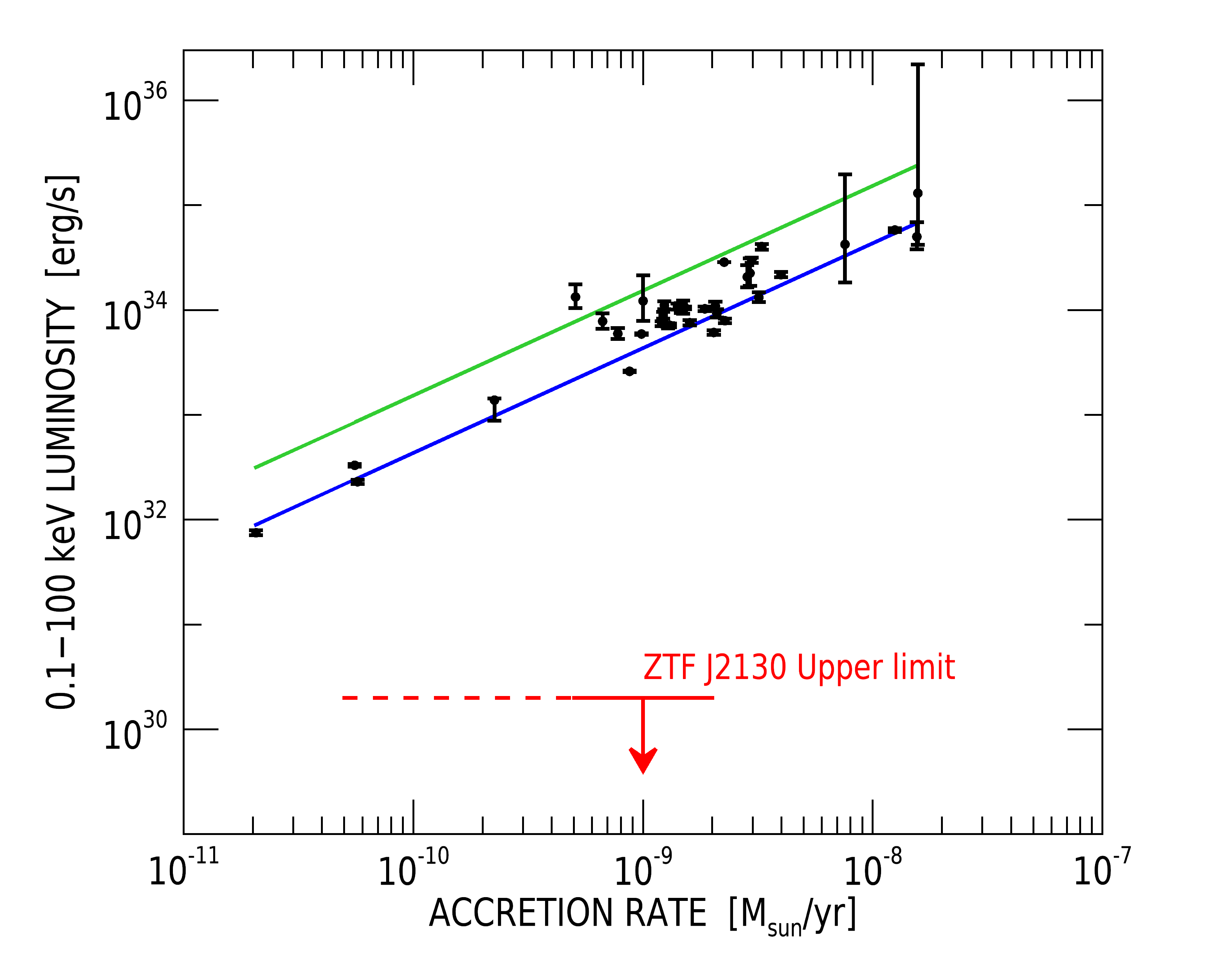}
   \caption{Luminosity versus mass accretion rate in a sample of intermediate polars studied by \citet{sul19}. The values predicted by eq. (1) are shown by the blue and green lines for  M$_{WD}$=1 and 0.5 $\msun$, respectively. The red line shows the upper limit we derived on the X-ray luminosity of \ztf . }
    \label{fig-IP}
\end{figure}

X-rays from accreting WDs can be emitted from the boundary layer between the disk and the star surface or,  in the case of intermediate polars, where  the magnetic field is sufficiently strong to channel the accretion flow, from  the shocked plasma in the accretion column. In this case, the maximum plasma temperature $T$ in the post shock region is an increasing function of the WD mass. A simple estimate  (e.g. \citealt{fra02}) is

\begin{equation}
T =   \frac{3}{8} \frac{G M \mu m_p}{k R },
\end{equation}
  
\noindent
which gives kT = 19  keV for \ztf\  
(m$_p$ is the proton mass and we used 
a mean molecular weight $\mu$ =0.615).
Our luminosity upper limit for a thermal bremsstrahlung spectrum of this temperature is compared in Fig.~\ref{fig-IP} with the luminosities  of a sample of intermediate polars observed in the hard X-ray range \citep{sul19}. It is clear that, even if the mass accretion rate derived by \citet{kup20} were overestimated by up to a factor $\sim$100, the \xmm\ limit implies a   luminosity much lower than that of intermediate polars, that disfavors the presence of a magnetic WD in this system.

If the WD in \ztf\ has a low magnetic field, the  available accretion power is mainly released in the boundary layer between the disk and the star, which might   be optically thick, and from the heated WD surface. The resulting  spectrum would peak at  soft X-ray/EUV energies  and  thus escape detection in the energy range covered by the  EPIC instrument. 
Considering the high inclination of this system, it is also possible that the  emitting region is occulted by the accretion disk. In this respect, it is interesting to note the anticorrelation between soft X-ray luminosity and inclination found in non-magnetic cataclysmic variables by \citet{van96}.  

Finally, we note the possibility that the  accretion rate derived   from the  evolutionary model used by \citet{kup20} has been overestimated.  An evolutionary phase with accretion rate at $\sim10^{-9}$ $\msun$ yr$^{-1}$   is required to explain the current properties of this system, but it is possible that this phase is now ending and the sdOB has already started to  shrink and underfill the Roche lobe.

\section{Conclusions} \label{sect:conclusions}

\ztf\ is the WD with a hot subdwarf binary companion with the shortest known orbital  period. The evidence, derived from optical observations, for an accretion disk around the WD, caused by Roche-lobe overflow of its sdB companion, motivated a search for X-ray emission. 

Our \xmm\  observation, despite reaching a flux limit more than two orders of magnitude below the previously available one, could not reveal X-ray emission. The luminosity upper limit is in the range $\sim$(0.5--2.5)$\times10^{30}$ erg s$^{-1}$,  depending on the assumed spectral  temperature. For the hard Bremsstrahlung spectra typical of accreting magnetic WDs, this limit is significantly below the luminosity expected from the mass accretion rate of  $\sim10^{-9}$ $\msun$ yr$^{-1}$ deduced from the evolution model of this binary.   
This suggests that   \ztf\  contains a non-magnetic WD,  that can have a luminosity up to a few $10^{33}$ erg s$^{-1}$, consistent with the theoretical accretion rate,  but characterized by a soft spectrum peaking in the EUV range and thus undetectable by  X-ray observatories. Alternative explanations for the lack of observable X-rays are either intrinsic absorption in the system, due to its  high inclination ($i$=86$^{\circ}$), or an overestimate of the current accretion rate implying that the Roche-lobe filling phase is nearly  ending.


\acknowledgments 
We acknowledge  support via ASI/INAF Agreement n. 2019-35-HH and PRIN-MIUR 2017 UnIAM (Unifying Isolated and Accreting Magnetars, PI S.~Mereghetti). 
TK acknowledges support from the National Science Foundation through grant AST \#2107982, from NASA through grant 80NSSC22K0338 and from STScI through grant HST-GO-16659.002-A. 
TRM acknowledges support from the UK's Science and Technology Facilities Council (STFC), grant number ST/T000406/1.
This work is based on data obtained with {\it XMM-Newton}, an ESA science mission with instruments and contributions directly funded by ESA Member States and the USA (NASA). 
This research has made use of the SIMBAD database, operated at CDS, Strasbourg, France.



\bibliographystyle{aasjournal} 





\begin{figure}
    \centering
      \includegraphics[width=0.7\textwidth]{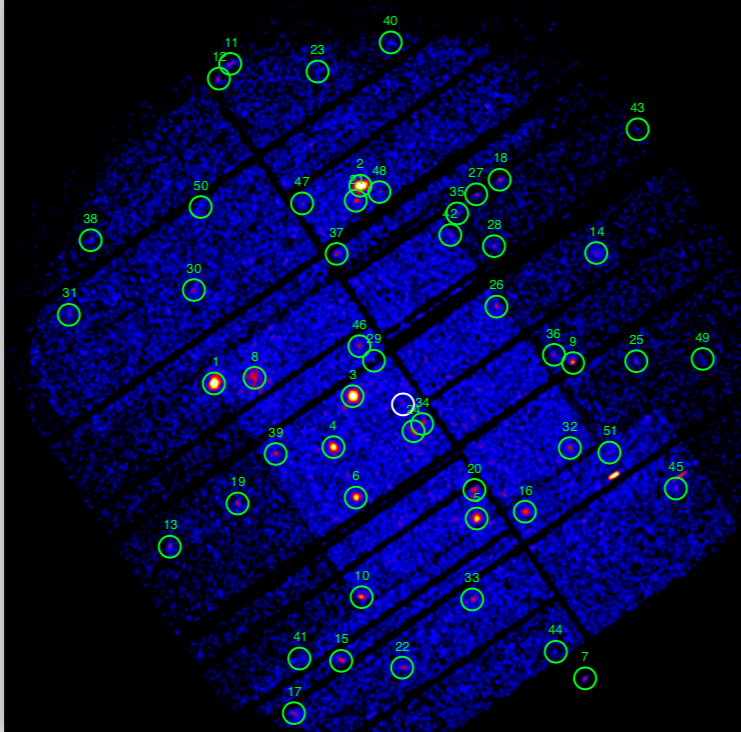}
 \caption{EPIC image in the 0.2-12 keV energy range of the sky region around \ztfl\ with the position of the 51 detected X-ray sources. North is to the top, east to the left. The white circle indicates the position of \ztf .  }
    \label{figure-sources}
\end{figure}

\appendix
We performed the detection, characterization, and identification of the several X-ray sources present in the EPIC images of the sky region around ZTF. To this aim, we applied the SAS task edetect\_chain, with the default configuration and input parameters, using the images in five energy ranges (0.2-0.5, 0.5-1, 1-2, 2-4.5, and 4.5-12 keV)  of the three EPIC cameras simultaneously. This resulted in the detection of 51 distinct sources (Fig.~\ref{figure-sources}). For each of them we obtained the X-ray position and its uncertainty, the detection likelihood (defined as L = -lnP, where P is the probability of a spurious detection due to a Poissonian random fluctuation of the background) and the determination of its point or extended nature. We derived the source fluxes from the measured count rates assuming the same absorbed power-law model (with photon index $\Gamma$ = 1.7 and hydrogen column density \nh = 3$\times 10^{20}$ cm$^{-2}$) used in the 3XMM source catalogue \citep{ros16}.

We used the task eposcorr to correct the position of the detected sources, through their cross-correlation with the Gaia EDR3 source catalogue \citep{GaiaCollaboration+16,GaiaCollaboration+21}. The average RA and DEC correction offset applied to the X-ray positions was 2.26 and 1.60 arcsec, respectively.

The last step of our analysis was the identification of each detected source with a known optical counterpart, or at least the assessment of its stellar or non-stellar nature. We performed this task with the cross-correlation with two different optical catalogues. The first choice was again the Gaia EDR3 catalogue, which to date is the most complete and accurate optical catalogue. For each X-ray source, we looked for all Gaia sources closer than three times the uncertainty of the X-ray position: due to the high accuracy of the Gaia positions, this approach gives us high confidence that no potential Gaia counterpart is missed. In this way we found at least one candidate Gaia counterpart for only 29 of the  sources.
Therefore, we integrated our analysis by cross-correlating the X-ray sources (with the same criteria) also with the Guide Star Catalogue (GSC, \citet{Lasker+08})\footnote{http://gsss.stsci.edu/webservices/GSC2/WebForm.aspx}, which    also provides a tentative classification of the source  nature\footnote{http://gsss.stsci.edu/Catalogs/GSC/GSC2/gsc23/gsc23\_release\_notes.htm\#ClassificationCodes}. We found a candidate GSC counterpart for   11 of the X-ray sources without Gaia counterparts. For 12 X-ray sources we found two or more candidate counterparts, while 11 X-ray sources remained uncorrelated to any optical counterpart.

In Table~\ref{table-sources} we report the main parameters of the detected X-ray sources and of their candidate counterparts, together with the proposed identification.
For each possible counterpart, we report the distance from the X-ray source, the source class given in the GSC catalogue (Star, Non-Star, or Unclassified), the magnitude type and value, and the estimated value and error (in logarithm) of the X-ray-to-optical flux ratio. 
Three of the brightest sources (\#10, \#13, and \#17) can unambiguously be identified with a known bright optical source of the Tycho catalogue \citep{Hog+00}. 

Two  sources  (\#8 and \#14) are clearly  extended. The analysis of the EPIC images shows that they are clearly visible between 0.5 and $\simeq$ 5 keV, while outside this energy range they are   confused with the background signal. Since for both objects there is no clear identification, we investigated their spectrum in order to obtain additional information on their properties. In both cases the count rate is very low (2.7$\times 10^{-2}$ and 7.8$\times 10^{-3}$ cts s$^{-1}$, respectively), so the count statistics is limited. The spectrum of source \#8 is rather hard and extends up to 10 keV. It can be equally well described with a power-law model with a photon index $\Gamma \simeq$ 1.6, or with a collisionally-heated plasma emission model with a high temperature (kT $\sim$ 10 keV). With both models the estimated flux (in the energy range 0.2-10 keV) is $\simeq 10^{-13}$ erg cm$^{-2}$ s$^{-1}$.  The high temperature of the thermal model suggests that the identification with a cluster of galaxies is unlikely. The hard power-law spectrum suggests a pulsar wind nebula nature, but no pulsars at this positions are known.  The spectrum of source \#14 is significantly softer: it can be described with either a power-law model with $\Gamma \simeq$ 2 or with a plasma model with kT $\sim$ 5 keV. Therefore, in this case the estimated plasma temperature is consistent with a cluster of galaxies or with a supernova remnant. However, the limited count statistics prevents to clearly discriminate between a thermal or non-thermal emission model. The estimated flux of this source is $\simeq 6\times 10^{-14}$ erg cm$^{-2}$ s$^{-1}$. In summary, based on the EPIC data, we cannot suggest any identification of these extended sources with a known class of X-ray sources.

\begin{longrotatetable}
\begin{deluxetable}{cccccccccccccl}
\tablecaption{Main parameters of the 51 detected X-ray sources. The sources are sorted by decreasing detection likelihood.}
\label{table-sources}
\tabletypesize{\scriptsize}
\startdata
(1)	 &	(2)			 &	(3)				 &	(4)	 &	(5)	 &		(6)							&	(7)			 &	(8)		 &	(9)	&	(10)	&	(11)		&	(12)					 &	(13)		 &	(14)	\\
N	 & 	RA$_{\rm X}$ & 	DEC$_{\rm X}$	 & 	ERR	 & 	CR	 & 		Flux ($\times 10^{-14}$	 	& 	Detection	 & 	Extended & 	$d$	&	Class	&	\multicolumn{2}{c}{Magnitude}			 & 	X/O	 ratio	 & 	Counterpart	\\	
n/a	 & 	$^\circ$	 & 	$^\circ$ & 	"	 & 	cts s$^{-1}$	 & 		erg cm$^{-2}$ s$^{-1}$)	 	& 	likelihood	 & 	source	 & 	"	&					n/a	&	Type	 & 	Value			 & 	Log10	 & 		\\	\hline
1	 & 	322.9037	 & 	44.3588	 & 	0.2	 & 	0.119	 $\pm$	0.004	 & 		12.50	 $\pm$		0.38	 & 	3343.42	 & 	NO	 & 	0.5	&					Star	&	GaiaG	 & 	11.2			 & 	-3.62	 $\pm$	0.05	 & 	Gaia EDR3 1970952112669846528	\\
	 &				 &			 &		 &							 &									 &			 &		 &		&							&			 &					 &							 &	1RXS J213137.1+442125	\\
2	 & 	322.7730	 & 	44.4847	 & 	0.2	 & 	0.141	 $\pm$	0.004	 & 		22.50	 $\pm$		0.58	 & 	3254.26	 & 	NO	 & 	0.4	&					Star	&	GaiaG	 & 	14.4			 & 	-2.07	 $\pm$	0.04	 & 	Gaia EDR3 1970957197911088256	\\	
3	 & 	322.7798	 & 	44.3510	 & 	0.2	 & 	0.061	 $\pm$	0.002	 & 		6.25	 $\pm$		0.18	 & 	2269.45	 & 	NO	 & 	0.2	&					-	&	GaiaG	 & 	14.3			 & 	-2.68	 $\pm$	0.01	 & 	Gaia EDR3 1970951219316446848	\\	
4	 & 	322.7973	 & 	44.3185	 & 	0.4	 & 	0.024	 $\pm$	0.001	 & 		7.11	 $\pm$		0.37	 & 	537.44	 & 	NO	 & 	0.6	&					Non-Star	&	SDSS\_r	 & 	22.4			 & 	1.38	 $\pm$	0.14	 & 	GSC2 N0331212284370	\\	
5	 & 	322.6702	 & 	44.2730	 & 	0.4	 & 	0.022	 $\pm$	0.001	 & 		7.70	 $\pm$		0.46	 & 	400.99	 & 	NO	 & 	-	&					-	&	-	 & 	-			 & 	-	  & 	?	\\	
6	 & 	322.7772	 & 	44.2866	 & 	0.4	 & 	0.019	 $\pm$	0.001	 & 		4.79	 $\pm$		0.30	 & 	301.96	 & 	NO	 & 	0.5	&					Non-Star	&	SDSS\_g	 & 	21.7			 & 	0.70	 $\pm$	0.10	 & 	GSC2 N0331212271069	\\	
7	 & 	322.5740	 & 	44.1716	 & 	0.6	 & 	0.016	 $\pm$	0.002	 & 		9.42	 $\pm$		0.90	 & 	199.27	 & 	NO	 & 	-	&					-	&	-	 & 	-			 & 	-	 	 & 	?	\\	
8	 & 	322.8676	 & 	44.3626	 & 	1.0	 & 	0.073	 $\pm$	0.005	 & 		14.20	 $\pm$		1.05	 & 	194.75	 & 	YES	 & 	2.9	&					Non-Star	&	SDSS\_r	 & 	21.4			 & 	1.30	 $\pm$	0.15	 & 	GSC2 N0331212285756	\\	
9	 & 	322.5840	 & 	44.3718	 & 	0.5	 & 	0.023	 $\pm$	0.002	 & 		4.96	 $\pm$		0.49	 & 	183.53	 & 	NO	 & 	-	&					-	&	-	 & 	-			 & 	-	 	 & 	?	\\	
10	 & 	322.7726	 & 	44.2233	 & 	0.6	 & 	0.014	 $\pm$	0.001	 & 		1.33	 $\pm$		0.11	 & 	171.66	 & 	NO	 & 	1.2	&					Star	&	GaiaG	 & 	10.2			 & 	-4.98	 $\pm$	0.11	 & 	TYC 3195-1291-1 	\\	
	 &				 &			 &		 &							 &									 &			 &		 &		&							&			 &					 &							 &	(Gaia EDR3 1970937647223334784)	\\
11	 & 	322.8890	 & 	44.5620	 & 	0.8	 & 	0.042	 $\pm$	0.005	 & 		23.70	 $\pm$		2.90	 & 	154.79	 & 	NO	 & 	1.1	&					Star	&	GaiaG	 & 	14.9			 & 	-1.85	 $\pm$	0.14	 & 	Gaia EDR3 1971050931274339456	\\	
12	 & 	322.8995	 & 	44.5525	 & 	0.9	 & 	0.021	 $\pm$	0.005	 & 		7.74	 $\pm$		1.92	 & 	112.35	 & 	NO	 & 	2.7	&					Star	&	GaiaG	 & 	18.6			 & 	-0.87	 $\pm$	0.44	 & 	Gaia EDR3 1971050720812613632	\\	
13	 & 	322.9422	 & 	44.2552	 & 	0.9	 & 	0.011	 $\pm$	0.001	 & 		0.98	 $\pm$		0.12	 & 	94.56	 & 	NO	 & 	0.9	&					Non-Star	&	SDSS\_i	 & 	14.7			 & 	-2.44	 $\pm$	0.29	 & 	GSC2 N0331212310632	\\	
	 & 		 & 		 & 		 & 				 & 						 & 		 & 		 & 	1.4	&					Star	&	GaiaG	 & 	11.4			 & 	-4.62	 $\pm$	0.29	 & 	TYC 3195-1009-1 	\\
	 &				 &			 &		 &							 &									 &			 &		 &		&							&			 &					 &							 &	(Gaia EDR3 1970938781091225472)	\\
	 & 		 & 		 & 		 & 				 & 						 & 		 & 		 & 	2.0	&					Non-Star	&	SDSS\_g	 & 	13.2			 & 	-3.39	 $\pm$	0.29	 & 	GSC2 N0331212271457	\\	
14	 & 	322.5631	 & 	44.4418	 & 	0.7	 & 	0.045	 $\pm$	0.005	 & 		9.83	 $\pm$		1.03	 & 	91.29	 & 	YES	 & 	-	&					-	&	-	 & 	-			 & 	-	 	 & 	?	\\	
15	 & 	322.7897	 & 	44.1830	 & 	0.8	 & 	0.013	 $\pm$	0.002	 & 		3.93	 $\pm$		0.51	 & 	75.79	 & 	NO	 & 	0.7	&					Star	&	GaiaG	 & 	19.8			 & 	-0.68	 $\pm$	0.20	 & 	Gaia EDR3 1970937372330018944	\\	
	 & 		 & 		 & 		 & 				 & 						 & 		 & 		 & 	1.1	&					-	&	GaiaG	 & 	17.1			 & 	-1.76	 $\pm$	0.06	 & 	Gaia EDR3 1970937372341684864	\\	
16	 & 	322.6268	 & 	44.2773	 & 	0.7	 & 	0.010	 $\pm$	0.001	 & 		5.43	 $\pm$		0.61	 & 	70.91	 & 	NO	 & 	1.4	&					Non-Star	&	SDSS\_i	 & 	21.8			 & 	1.15	 $\pm$	0.10	 & 	GSC2 N0331212276990	\\	
	 & 		 & 		 & 		 & 				 & 						 & 		 & 		 & 	2.0	&					Unclassified	&	WISE\_w1	 & 	15.7			 & 	-	 	 & 	GSC2 N0331212380540	\\	
17	 & 	322.8324	 & 	44.1495	 & 	0.8	 & 	0.017	 $\pm$	0.002	 & 		6.91	 $\pm$		0.90	 & 	70.69	 & 	NO	 & 	0.8	&					Non-Star	&	SDSS\_r	 & 	11.4			 & 	-3.03	 $\pm$	0.16	 & 	GSC2 N0331212325476	\\	
	 & 		 & 		 & 		 & 				 & 						 & 		 & 		 & 	1.3	&					Star	&	GaiaG	 & 	11.2			 & 	-3.89	 $\pm$	0.15	 & 	TYC 3195-1810-1	\\	
	 &				 &			 &		 &							 &									 &			 &		 &		&							&			 &					 &							 &	(Gaia EDR3 1970936375909299712)	\\
	 & 		 & 		 & 		 & 				 & 						 & 		 & 		 & 	2.2	&					Star	&	GaiaG	 & 	19.5			 & 	-0.56	 $\pm$	0.15	 & 	Gaia EDR3 1970936375899213440	\\	
18	 & 	322.6491	 & 	44.4884	 & 	1.1	 & 	0.014	 $\pm$	0.002	 & 		4.83	 $\pm$		0.65	 & 	66.51	 & 	NO	 & 	1.3	&					Non-Star	&	SDSS\_g	 & 	21.9			 & 	0.78	 $\pm$	0.20	 & 	GSC2 N0331212254829	\\	
19	 & 	322.8823	 & 	44.2828	 & 	0.9	 & 	0.009	 $\pm$	0.001	 & 		2.60	 $\pm$		0.35	 & 	62.60	 & 	NO	 & 	1.1	&					Star	&	GaiaG	 & 	16.8			 & 	-2.07	 $\pm$	0.20	 & 	Gaia EDR3 1970939537005448064	\\	
20	 & 	322.6726	 & 	44.2914	 & 	0.9	 & 	0.008	 $\pm$	0.001	 & 		0.76	 $\pm$		0.09	 & 	54.98	 & 	NO	 & 	2.1	&					Star	&	GaiaG	 & 	16.3			 & 	-2.80	 $\pm$	0.27	 & 	Gaia EDR3 1970944931484039552	\\	
21	 & 	322.7768	 & 	44.4750	 & 	0.9	 & 	0.010	 $\pm$	0.001	 & 		1.44	 $\pm$		0.21	 & 	47.15	 & 	NO	 & 	0.5	&					Star	&	GaiaG	 & 	18.8			 & 	-1.52	 $\pm$	0.36	 & 	Gaia EDR3 1970956437691046144	\\	
22	 & 	322.7357	 & 	44.1784	 & 	0.9	 & 	0.014	 $\pm$	0.002	 & 		2.25	 $\pm$		0.36	 & 	44.78	 & 	NO	 & 	-	&					-	&	-	 & 	-			 & 	-	 	 & 	?	\\	
23	 & 	322.8111	 & 	44.5572	 & 	1.3	 & 	0.011	 $\pm$	0.002	 & 		0.59	 $\pm$		0.13	 & 	37.34	 & 	NO	 & 	2.7	&					Non-Star	&	SDSS\_r	 & 	21.7			 & 	0.01	 $\pm$	0.96	 & 	GSC2 N0331212239468	\\	
	 & 		 & 		 & 		 & 				 & 						 & 		 & 		 & 	3.3	&					Non-Star	&	SDSS\_r	 & 	21.6			 & 	-0.03	 $\pm$	0.96	 & 	GSC2 N0331212254752	\\	
	 & 		 & 		 & 		 & 				 & 						 & 		 & 		 & 	3.3	&					Non-Star	&	SDSS\_i	 & 	21.2			 & 	-0.05	 $\pm$	0.96	 & 	GSC2 N0331212265386	\\	
24	 & 	322.7257	 & 	44.3285	 & 	1.0	 & 	0.006	 $\pm$	0.001	 & 		2.10	 $\pm$		0.28	 & 	37.10	 & 	NO	 & 	-	&					-	&	-	 & 	-			 & 	-	 	 & 	?	\\	
25	 & 	322.5279	 & 	44.3729	 & 	1.1	 & 	0.008	 $\pm$	0.001	 & 		1.98	 $\pm$		0.32	 & 	34.78	 & 	NO	 & 	2.0	&					Star	&	GaiaG	 & 	17.9			 & 	-1.72	 $\pm$	0.26	 & 	Gaia EDR3 1970996157547913856	\\	
26	 & 	322.6523	 & 	44.4078	 & 	0.8	 & 	0.005	 $\pm$	0.001	 & 		0.75	 $\pm$		0.12	 & 	34.74	 & 	NO	 & 	-	&					-	&	-	 & 	-			 & 	-	 	 & 	?	\\	
27	 & 	322.6700	 & 	44.4790	 & 	1.1	 & 	0.008	 $\pm$	0.001	 & 		1.13	 $\pm$		0.19	 & 	34.26	 & 	NO	 & 	0.7	&					Star	&	GaiaG	 & 	14.8			 & 	-3.22	 $\pm$	0.36	 & 	Gaia EDR3 1971002827645828352	\\	
28	 & 	322.6544	 & 	44.4462	 & 	1.0	 & 	0.007	 $\pm$	0.001	 & 		1.12	 $\pm$		0.22	 & 	30.88	 & 	NO	 & 	2.2	&					Non-Star	&	SDSS\_g	 & 	21.9			 & 	0.14	 $\pm$	0.61	 & 	GSC2 N0331212233476	\\	
29	 & 	322.7616	 & 	44.3734	 & 	0.9	 & 	0.007	 $\pm$	0.003	 & 		2.82	 $\pm$		0.98	 & 	29.88	 & 	NO	 & 	0.9	&					Non-Star	&	SDSS\_i	 & 	22.0			 & 	0.94	 $\pm$	0.39	 & 	GSC2 N0331212287517	\\	
30	 & 	322.9213	 & 	44.4183	 & 	1.0	 & 	0.009	 $\pm$	0.002	 & 		2.52	 $\pm$		0.45	 & 	29.64	 & 	NO	 & 	2.7	&					Star	&	GaiaG	 & 	18.2			 & 	-1.53	 $\pm$	0.31	 & 	Gaia EDR3 1970952524975861504	\\	
	 & 		 & 		 & 		 & 				 & 						 & 		 & 		 & 	3.0	&					Star	&	GaiaG	 & 	13.3			 & 	-3.48	 $\pm$	0.31	 & 	Gaia EDR3 1970952524986698624	\\	
31	 & 	323.0323	 & 	44.4024	 & 	1.3	 & 	0.013	 $\pm$	0.002	 & 		1.29	 $\pm$		0.20	 & 	29.19	 & 	NO	 & 	2.5	&					Star	&	GaiaG	 & 	16.2			 & 	-2.62	 $\pm$	0.38	 & 	Gaia EDR3 1971034266801276288	\\	
32	 & 	322.5873	 & 	44.3178	 & 	1.0	 & 	0.006	 $\pm$	0.001	 & 		2.34	 $\pm$		0.38	 & 	28.27	 & 	NO	 & 	1.6	&					Non-Star	&	SDSS\_i	 & 	21.1			 & 	0.49	 $\pm$	0.20	 & 	GSC2 N0331212281960	\\	
33	 & 	322.6738	 & 	44.2218	 & 	1.0	 & 	0.007	 $\pm$	0.001	 & 		1.10	 $\pm$		0.17	 & 	27.42	 & 	NO	 & 	2.3	&					Star	&	GaiaG	 & 	20.1			 & 	-1.13	 $\pm$	0.29	 & 	Gaia EDR3 1970944446140502272	\\	
34	 & 	322.7184	 & 	44.3332	 & 	0.9	 & 	0.004	 $\pm$	0.001	 & 		0.60	 $\pm$		0.09	 & 	25.85	 & 	NO	 & 	1.4	&					Star	&	GaiaG	 & 	17.3			 & 	-2.48	 $\pm$	0.24	 & 	Gaia EDR3 1970950978798273920	\\	
35	 & 	322.6870	 & 	44.4670	 & 	1.1	 & 	0.008	 $\pm$	0.001	 & 		4.28	 $\pm$		0.78	 & 	23.26	 & 	NO	 & 	2.9	&					Non-Star	&	SDSS\_r	 & 	21.6			 & 	0.85	 $\pm$	0.17	 & 	GSC2 N0331212262403	\\	
36	 & 	322.6012	 & 	44.3770	 & 	1.2	 & 	0.006	 $\pm$	0.001	 & 		3.43	 $\pm$		0.61	 & 	22.24	 & 	NO	 & 	1.6	&					Star	&	GaiaG	 & 	20.9			 & 	-0.29	 $\pm$	0.16	 & 	Gaia EDR3 1970949260801367296	\\	
37	 & 	322.7944	 & 	44.4414	 & 	1.1	 & 	0.006	 $\pm$	0.001	 & 		1.95	 $\pm$		0.37	 & 	22.00	 & 	NO	 & 	1.7	&					Star	&	GaiaG	 & 	15.0			 & 	-2.89	 $\pm$	0.30	 & 	Gaia EDR3 1970956167118744320	\\	
38	 & 	323.0129	 & 	44.4497	 & 	1.0	 & 	0.015	 $\pm$	0.003	 & 		11.40	 $\pm$		2.40	 & 	21.17	 & 	NO	 & 	2.9	&					Star	&	GaiaG	 & 	20.3			 & 	-0.03	 $\pm$	0.09	 & 	Gaia EDR3 1971046292698993152	\\	
	 & 		 & 		 & 		 & 				 & 						 & 		 & 		 & 	3.1	&					Star	&	GaiaG	 & 	17.9			 & 	-1.00	 $\pm$	0.17	 & 	Gaia EDR3 1971046288408505344	\\	
39	 & 	322.8485	 & 	44.3143	 & 	1.0	 & 	0.006	 $\pm$	0.001	 & 		1.93	 $\pm$		0.31	 & 	20.26	 & 	NO	 & 	1.8	&					Non-Star	&	SDSS\_g	 & 	22.2			 & 	0.48	 $\pm$	0.20	 & 	GSC2 N0331212276215	\\	
40	 & 	322.7466	 & 	44.5755	 & 	2.3	 & 	0.011	 $\pm$	0.004	 & 		7.22	 $\pm$		2.36	 & 	17.86	 & 	NO	 & 	2.2	&					Star	&	GaiaG	 & 	15.3			 & 	-2.24	 $\pm$	0.35	 & 	Gaia EDR3 1971004957949785728	\\	
	 & 		 & 		 & 		 & 				 & 						 & 		 & 		 & 	2.6	&					Non-Star	&	SDSS\_g	 & 	20.7			 & 	0.48	 $\pm$	0.35	 & 	GSC2 N0331212232458	\\	
41	 & 	322.8268	 & 	44.1844	 & 	1.4	 & 	0.003	 $\pm$	0.001	 & 		1.51	 $\pm$		0.72	 & 	16.89	 & 	NO	 & 	2.2	&					Star	&	GaiaG	 & 	20.1			 & 	-0.97	 $\pm$	0.50	 & 	Gaia EDR3 1970936650776371968	\\	
	 & 		 & 		 & 		 & 				 & 						 & 		 & 		 & 	4.2	&					Star	&	GaiaG	 & 	20.0			 & 	-1.04	 $\pm$	0.50	 & 	Gaia EDR3 1970936650787455744	\\	
42	 & 	322.6930	 & 	44.4533	 & 	1.1	 & 	0.005	 $\pm$	0.001	 & 		1.70	 $\pm$		0.37	 & 	16.48	 & 	NO	 & 	-	&					-	&	-	 & 	-			 & 	-	 	 & 	?	\\	
43	 & 	322.5266	 & 	44.5201	 & 	1.0	 & 	0.011	 $\pm$	0.006	 & 		8.61	 $\pm$		4.49	 & 	16.15	 & 	NO	 & 	2.1	&					Non-Star	&	GaiaG	 & 	21.1			 & 	0.16	 $\pm$	0.23	 & 	Gaia EDR3 1971006572852951296	\\	
44	 & 	322.6004	 & 	44.1886	 & 	1.3	 & 	0.004	 $\pm$	0.001	 & 		0.40	 $\pm$		0.14	 & 	15.54	 & 	NO	 & 	-	&					-	&	-	 & 	-			 & 	-	 	 & 	?	\\	
45	 & 	322.4930	 & 	44.2919	 & 	1.1	 & 	0.007	 $\pm$	0.001	 & 		1.70	 $\pm$		0.35	 & 	15.06	 & 	NO	 & 	-	&					-	&	-	 & 	-			 & 	-	 	 & 	?	\\	
46	 & 	322.7744	 & 	44.3828	 & 	1.0	 & 	0.004	 $\pm$	0.001	 & 		1.34	 $\pm$		0.24	 & 	15.00	 & 	NO	 & 	0.6	&					-	&	GaiaG	 & 	17.7			 & 	-1.99	 $\pm$	0.08	 & 	Gaia EDR3 1970951528542823296	\\	
	 & 		 & 		 & 		 & 				 & 						 & 		 & 		 & 	1.1	&					Star	&	GaiaG	 & 	20.1			 & 	-1.04	 $\pm$	0.21	 & 	Gaia EDR3 1970951528554080768	\\	
	 & 		 & 		 & 		 & 				 & 						 & 		 & 		 & 	2.5	&					Star	&	GaiaG	 & 	20.9			 & 	-0.73	 $\pm$	0.21	 & 	Gaia EDR3 1970951528543900544	\\	
	 & 		 & 		 & 		 & 				 & 						 & 		 & 		 & 	2.5	&					Non-Star	&	SDSS\_z	 & 	19.8			 & 	-0.18	 $\pm$	0.21	 & 	GSC2 N0331212259744	\\	
47	 & 	322.8249	 & 	44.4733	 & 	1.3	 & 	0.006	 $\pm$	0.002	 & 		4.47	 $\pm$		1.05	 & 	13.65	 & 	NO	 & 	-	&					-	&	-	 & 	-			 & 	-	 	 & 	?	\\	
48	 & 	322.7561	 & 	44.4806	 & 	1.2	 & 	0.005	 $\pm$	0.001	 & 		0.35	 $\pm$		0.08	 & 	11.87	 & 	NO	 & 	2.3	&					Star	&	GaiaG	 & 	17.0			 & 	-2.87	 $\pm$	0.92	 & 	Gaia EDR3 1970957094831654784	\\	
	 & 		 & 		 & 		 & 				 & 						 & 		 & 		 & 	3.5	&					Star	&	GaiaG	 & 	19.4			 & 	-1.89	 $\pm$	0.92	 & 	Gaia EDR3 1970957094820313856	\\	
49	 & 	322.4694	 & 	44.3742	 & 	2.0	 & 	0.006	 $\pm$	0.001	 & 		0.87	 $\pm$		0.20	 & 	10.76	 & 	NO	 & 	1.6	&					Star	&	GaiaG	 & 	18.9			 & 	-1.68	 $\pm$	0.38	 & 	Gaia EDR3 1970996672943997568	\\	
50	 & 	322.9151	 & 	44.4709	 & 	2.4	 & 	0.008	 $\pm$	0.002	 & 		2.14	 $\pm$		0.50	 & 	10.68	 & 	NO	 & 	3.8	&					Non-Star	&	SDSS\_z	 & 	20.5			 & 	0.30	 $\pm$	0.37	 & 	GSC2 N0331212267844	\\	
	 & 		 & 		 & 		 & 				 & 						 & 		 & 		 & 	6.2	&					Star	&	GaiaG	 & 	18.3			 & 	-1.55	 $\pm$	0.37	 & 	Gaia EDR3 1971047284838488704	\\	
	 & 		 & 		 & 		 & 				 & 						 & 		 & 		 & 	6.9	&					Non-Star	&	SDSS\_i	 & 	20.6			 & 	0.26	 $\pm$	0.37	 & 	GSC2 N0331212251938	\\	
51	 & 	322.5518	 & 	44.3149	 & 	1.5	 & 	0.006	 $\pm$	0.002	 & 		3.45	 $\pm$		0.93	 & 	10.03	 & 	NO	 & 	3.0	&					Star	&	GaiaG	 & 	16.3			 & 	-2.13	 $\pm$	0.18	 & 	Gaia EDR3 1970948504896789632	\\	\hline
\enddata
\tablecomments{Column description: (1) X-ray source ID number; (2), (3) X-ray position; (4) X-ray position uncertainty; (5) count rate in the range 0.2-12 keV; (6) flux in the range 0.2-12 keV, assuming an absorbed power-law spectrum with photon index $\Gamma$ = 1.7 and hydrogen column density \nh = 3$\times 10^{20}$ cm$^{-2}$; (7) detection likelihood; (8) source morphology; (9)  angular distance of the candidate optical counterpart from the X-ray position; (10) classification of the candidate counterpart; (11) magnitude type; (12) magnitude value; (13) logarithmic value of the X-ray-to-optical flux ratio; (14) proposed source identification.}
\end{deluxetable}
\end{longrotatetable}

\end{document}